\documentclass[twocolumn]{aastex63}

\usepackage{amsmath}

\graphicspath{{./}}

\received{\today}
\revised{}
\accepted{22 Apr 2020}
\submitjournal{ApJL}

\shorttitle{Iron from the KELT-9b day-side atmosphere}
\shortauthors{Pino et al.}
\usepackage{natbib}
\bibpunct{(}{)}{;}{a}{}{,}

\begin{document}

\title{Neutral Iron Emission Lines From The Day-side Of KELT-9b\\ The GAPS Programme With HARPS-N At TNG XX}

\correspondingauthor{Lorenzo Pino}
\email{l.pino@uva.nl}

\author[0000-0002-1321-8856]{Lorenzo Pino}
\affil{Anton Pannekoek Institute for Astronomy, University of Amsterdam \\
Science Park 904 \\
1098 XH Amsterdam, The Netherlands}

\author{Jean-Michel D\'esert}
\affiliation{Anton Pannekoek Institute for Astronomy, University of Amsterdam \\
Science Park 904 \\
1098 XH Amsterdam, The Netherlands}

\author{Matteo Brogi}
\affiliation{Department of Physics, University of Warwick\\
Coventry CV4 7AL, UK}
\affiliation{INAF - Osservatorio Astrofisico di Torino\\
Via Osservatorio 20\\
I-10025, Pino Torinese, Italy}
\affiliation{Centre for Exoplanets and Habitability, University of Warwick\\
Gibbet Hill Road\\
Coventry CV4 7AL, UK}

\author{Luca Malavolta}
\affiliation{Dipartimento di Fisica e Astronomia ‘Galileo Galilei’, Universit\`a di Padova\\
Vicolo dell’Osservatorio 3\\
I-35122, Padova, Italy}
\affiliation{INAF - Osservatorio Astrofisico di Catania\\
Via S. Sofia 78\\
I-95123, Catania, Italy}

\author{Aur\'elien Wyttenbach}
\affiliation{Leiden Observatory, Leiden University\\
Postbus 9513\\
2300 RA Leiden, The Netherlands}

\author{Michael Line}
\affiliation{Arizona State University\\
Tempe, AZ, USA}

\author{Jens Hoeijmakers}
\affiliation{University of Bern, Center for Space and Habitability\\
Gesellschaftsstrasse 6\\
 CH-3012, Bern, Switzerland}
 \affiliation{Observatoire astronomique de l'Universit\'e de G\`eneve\\
 51 Chemin des Maillettes\\
 1290 Versoix, Switzerland.}

\author{Luca Fossati}
\affiliation{Space Research Institute, Austrian Academy of Sciences\\
 Schmiedlstrasse 6\\
 A-8041 Graz, Austria}

\author{Aldo Stefano Bonomo}
\affiliation{INAF - Osservatorio Astrofisico di Torino\\
Via Osservatorio 20\\
I-10025, Pino Torinese, Italy}
 
\author{Valerio Nascimbeni}
\affiliation{INAF - Osservatorio Astronomico di Padova\\
Vicolo dell’Osservatorio 5\\
35122 Padova, Italy} 
 
 \author{Vatsal Panwar}
 \affiliation{Anton Pannekoek Institute for Astronomy, University of Amsterdam \\
Science Park 904 \\
1098 XH Amsterdam, The Netherlands}
 
\author{Laura Affer} 
\affiliation{INAF – Osservatorio Astronomico di Palermo\\
Piazza del Parlamento, 1\\
I-90134 Palermo, Italy}  

\author{Serena Benatti}
 \affiliation{INAF – Osservatorio Astronomico di Palermo\\
Piazza del Parlamento, 1\\
I-90134 Palermo, Italy} 

\author{Katia Biazzo}
\affiliation{INAF - Osservatorio Astrofisico di Catania\\
Via S. Sofia 78\\
I-95123, Catania, Italy}

\author{Andrea Bignamini}
\affiliation{INAF - Osservatorio Astronomico di Trieste\\
Via Tiepolo 11\\
34143 Trieste, Italy}

\author{Franscesco Borsa}
\affiliation{INAF - Osservatorio Astronomico di Brera\\
Via E. Bianchi 46\\
23807 Merate, Italy
}
 
\author{Ilaria Carleo}
\affiliation{Astronomy Department and Van Vleck Observatory\\
Wesleyan University\\
CT 06459 Middletown, USA}
\affiliation{INAF - Osservatorio Astronomico di Padova\\
Vicolo dell’Osservatorio 5\\
35122 Padova, Italy}

\author{Riccardo Claudi}
\affiliation{INAF - Osservatorio Astronomico di Padova\\
Vicolo dell’Osservatorio 5\\
35122 Padova, Italy}

\author{Rosario Cosentino}
\affiliation{Fundaci\'on Galileo Galilei-INAF \\
Rambla Jos\'e Ana Fernandez P\'erez 7 \\
38712 Bre\~na Baja, TF, Spain}

\author{Elvira Covino}
\affiliation{INAF - Osservatorio Astronomico di Capodimonte\\
Salita Moiariello 16\\
80131 Napoli, Italy}

\author{Mario Damasso}
\affiliation{INAF - Osservatorio Astrofisico di Torino\\
Via Osservatorio 20\\
I-10025, Pino Torinese, Italy}

\author{Silvano Desidera}
\affiliation{INAF - Osservatorio Astronomico di Padova\\
Vicolo dell’Osservatorio 5\\
35122 Padova, Italy}

\author{Paolo Giacobbe}
\affiliation{INAF - Osservatorio Astrofisico di Torino\\
Via Osservatorio 20\\
I-10025, Pino Torinese, Italy}

\author{Avet Harutyunyan}
\affiliation{Fundaci\'on Galileo Galilei-INAF \\
Rambla Jos\'e Ana Fernandez P\'erez 7 \\
38712 Bre\~na Baja, TF, Spain}

\author{Antonino Francesco Lanza}
\affiliation{INAF - Osservatorio Astrofisico di Catania\\
Via S. Sofia 78\\
I-95123, Catania, Italy}

\author{Giuseppe Leto}
\affiliation{INAF - Osservatorio Astrofisico di Catania\\
Via S. Sofia 78\\
I-95123, Catania, Italy}

\author{Antonio Maggio}
\affiliation{INAF – Osservatorio Astronomico di Palermo\\
Piazza del Parlamento, 1\\
I-90134 Palermo, Italy} 
 
\author{Jesus Maldonado} 
 \affiliation{INAF – Osservatorio Astronomico di Palermo\\
Piazza del Parlamento, 1\\
I-90134 Palermo, Italy}

\author{Luigi Mancini}
\affiliation{Department of Physics, University of Rome Tor Vergata\\
Via della Ricerca Scientifica 1\\
I-00133 Rome, Italy}
\affiliation{INAF - Osservatorio Astrofisico di Torino\\
Via Osservatorio 20\\
I-10025, Pino Torinese, Italy}
\affiliation{Max Planck Institute for Astronomy\\
Königstuhl 17\\
69117 Heidelberg, Germany}

\author{Giuseppina Micela}
\affiliation{INAF – Osservatorio Astronomico di Palermo\\
Piazza del Parlamento, 1\\
I-90134 Palermo, Italy}

\author{Emilio Molinari}
\affiliation{INAF Osservatorio Astronomico di Cagliari \& REM\\
Via della Scienza, 5\\ 
I-09047 Selargius CA, Italy}

\author{Isabella Pagano}
\affiliation{INAF - Osservatorio Astrofisico di Catania\\
Via S. Sofia 78\\
I-95123, Catania, Italy}

\author{Giampaolo Piotto}
\affiliation{Dipartimento di Fisica e Astronomia ‘Galileo Galilei’, Universit\`a di Padova\\
Vicolo dell’Osservatorio 3\\
I-35122, Padova, Italy}

\author{Ennio Poretti}
\affiliation{Fundaci\'on Galileo Galilei-INAF \\
Rambla Jos\'e Ana Fernandez P\'erez 7 \\
38712 Bre\~na Baja, TF, Spain}
\affiliation{INAF - Osservatorio Astronomico di Brera\\
Via E. Bianchi 46\\
23807 Merate, Italy
}

\author{Monica Rainer}
\affiliation{INAF-Osservatorio Astrofisico di Arcetri\\
Largo Enrico Fermi 5\\
I-50125 Firenze, Italy}

\author{Gaetano Scandariato}
\affiliation{INAF - Osservatorio Astrofisico di Catania\\
Via S. Sofia 78\\
I-95123, Catania, Italy}

\author{Alessandro Sozzetti}
\affiliation{INAF - Osservatorio Astrofisico di Torino\\
Via Osservatorio 20\\
I-10025, Pino Torinese, Italy}

 \author{Romain Allart}
 \affiliation{Observatoire astronomique de l'Universit\'e de G\`eneve\\
 51 Chemin des Maillettes\\
 1290 Versoix, Switzerland.}
 
\author{Luca Borsato}
\affiliation{Dipartimento di Fisica e Astronomia ‘Galileo Galilei’, Universit\`a di Padova\\
Vicolo dell’Osservatorio 3\\
I-35122, Padova, Italy}

\author{Giovanni Bruno}
\affiliation{INAF - Osservatorio Astrofisico di Catania\\
Via S. Sofia 78\\
I-95123, Catania, Italy}

\author{Luca Di Fabrizio}
\affiliation{Fundaci\'on Galileo Galilei-INAF \\
Rambla Jos\'e Ana Fernandez P\'erez 7 \\
38712 Bre\~na Baja, TF, Spain}

\author{David Ehrenreich}
 \affiliation{Observatoire astronomique de l'Universit\'e de G\`eneve\\
 51 Chemin des Maillettes\\
 1290 Versoix, Switzerland.}

\author{Aldo Fiorenzano}
\affiliation{Fundaci\'on Galileo Galilei-INAF \\
Rambla Jos\'e Ana Fernandez P\'erez 7 \\
38712 Bre\~na Baja, TF, Spain}
 
  \author{Giuseppe Frustagli}
\affiliation{INAF - Osservatorio Astronomico di Brera\\
Via E. Bianchi 46\\
23807 Merate, Italy
}
\affiliation{Dipartimento di Fisica G. Occhialini, Universita\`a degli Studi di Milano-Bicocca\\
Piazza della Scienza 3\\
20126 Milano, Italy}

\author{Baptiste Lavie}
 \affiliation{Observatoire astronomique de l'Universit\'e de G\`eneve\\
 51 Chemin des Maillettes\\
 1290 Versoix, Switzerland.}
 
 \author{Christophe Lovis}
  \affiliation{Observatoire astronomique de l'Universit\'e de G\`eneve\\
 51 Chemin des Maillettes\\
 1290 Versoix, Switzerland.}
 
\author{Antonio Magazz\`u}
\affiliation{Fundaci\'on Galileo Galilei-INAF \\
Rambla Jos\'e Ana Fernandez P\'erez 7 \\
38712 Bre\~na Baja, TF, Spain} 

\author{Domenico Nardiello}
\affiliation{Dipartimento di Fisica e Astronomia ‘Galileo Galilei’, Universit\`a di Padova\\
Vicolo dell’Osservatorio 3\\
I-35122, Padova, Italy}

\author{Marco Pedani}
\affiliation{Fundaci\'on Galileo Galilei-INAF \\
Rambla Jos\'e Ana Fernandez P\'erez 7 \\
38712 Bre\~na Baja, TF, Spain}

\author{Riccardo Smareglia}
\affiliation{INAF - Osservatorio Astronomico di Trieste\\
Via Tiepolo 11\\
34143 Trieste, Italy}



\begin{abstract}
We present the first detection of atomic emission lines from the atmosphere of an exoplanet. We detect neutral iron lines from the day-side of KELT-9b ($T_\mathrm{eq}\sim4,000~\mathrm{K}$). We combined thousands of spectrally resolved lines observed during one night with the HARPS-N spectrograph ($R\sim115,000$), mounted at the Telescopio Nazionale Galileo. We introduce a novel statistical approach to extract the planetary parameters from the binary mask cross-correlation analysis. We also adapt the concept of contribution function to the context of high spectral resolution observations, to identify the location in the planetary atmosphere where the detected emission originates. The average planetary line profile intersected by a stellar G2 binary mask was found in emission with a contrast of $84\pm14~\mathrm{ppm}$ relative to the planetary plus stellar continuum ($40\pm5\%$ relative to the planetary continuum only). This result unambiguously indicates the presence of an atmospheric thermal inversion. Finally, assuming a modelled temperature profile previously published \citep{Lothringer2018}, we show that an iron abundance consistent with a few times the stellar value explains the data well. In this scenario, the iron emission originates at the $10^{-3}$--$10^{-5}~\mathrm{bar}$ level.
\end{abstract}

\keywords{Exoplanet atmospheres --- Exoplanet atmospheric composition --- Hot Jupiters  --- High resolution spectroscopy}


\section{Introduction}
Ultra-hot Jupiters are tidally locked gaseous giant planets that orbit their host stars in mere hours or days, often reaching temperatures above $2,500~\mathrm{K}$ in their permanent day-sides. Unlike for their cooler counterparts, equilibrium chemistry should provide an accurate description of their chemical network, and known condensates are likely secluded to their night-side \citep{Kitzmann2018, Lothringer2018, Parmentier2018, Helling2019_w18}.\\
Detections of atomic metals at the day-night transition of their atmospheres (WASP-12b, \citealt{Fossati2010, Haswell2012}, KELT-9b, \citealt{Hoeijmakers2018_k9, Hoeijmakers2019, Cauley2019}; MASCARA-2b, \citealt{CasasayasBarris2019}; WASP-121b, \citealt{Sing2019, Gibson2020}) show that heavy elements are not necessarily sequestered deep in the atmosphere of these planets. This may also indicate the presence of a shallow radiative-convective boundary \citep{Thorngren2019}.\\
Iron is an element of particular interest. Indeed, its abundance is a proxy for the metallicity of stars, and thus a particularly relevant case for comparison between planetary and stellar metallicity. It was detected both in neutral and ionized form at the day-night transition in the atmosphere of KELT-9b \citep{Hoeijmakers2018_k9, Cauley2019, Borsa2019}, probing pressures as low as a few $\mu\mathrm{bar}$ \citep{Hoeijmakers2019}. These lines likely originate within the extended atmosphere of the planet, also detected with $\mathrm{H}\alpha$ and \ion{Ca}{2} lines \citep{Yan2018, Turner2020, Yan2019}. Ionized iron  was also found in the upper atmosphere of MASCARA-2b \citep{CasasayasBarris2019} and in the exospheres of WASP-12b and WASP-121b \citep{Haswell2012, Sing2019}. Yet, a detection of photospheric planetary iron lines is still missing.\\
In this paper we employ the high-resolution ($R\sim115,000$) spectrograph HARPS-N, mounted at the Telescopio Nazionale Galileo (TNG), to observe for the first time the thermal emission of an exoplanet with this instrument. To do so, we targeted KELT-9b monitoring the planet from quadrature to right before the planet is eclipsed behind the star. We describe our observations and data reduction in section \ref{sec:methods} and Appendix \ref{sec:appendix_pipeline}, including an adaptation of the line-weighted stellar binary masks method, traditionally used to extract radial velocities of exoplanets, to extract the signal of the planet via a cross-correlation function (CCF). We present the results of this analysis in Sec. \ref{sec:results}. We then perform a second analysis of the extracted planetary CCF to derive atmospheric parameters of the planet based on models (Sec. \ref{sec:methods_models}, Sec. \ref{sec:results_models}). To this aim, we introduce a new method to compare models and observations making use of the CCF technique with a line weighted binary mask, present a novel adaptation of the concept of contribution function to the context of cross-correlation analyses (Section \ref{sec:methods_models}), and apply these tools to our observations (Sec. \ref{sec:results_models}). We discuss the implications of our study in Section \ref{sec:discussion}.

\section{Methods: treatment of data}
\label{sec:methods}

\subsection{Observations and data reduction}
\label{sec:methods:observations_and_data_reduction}
\begin{figure*}
\centering
\includegraphics[width=17cm]{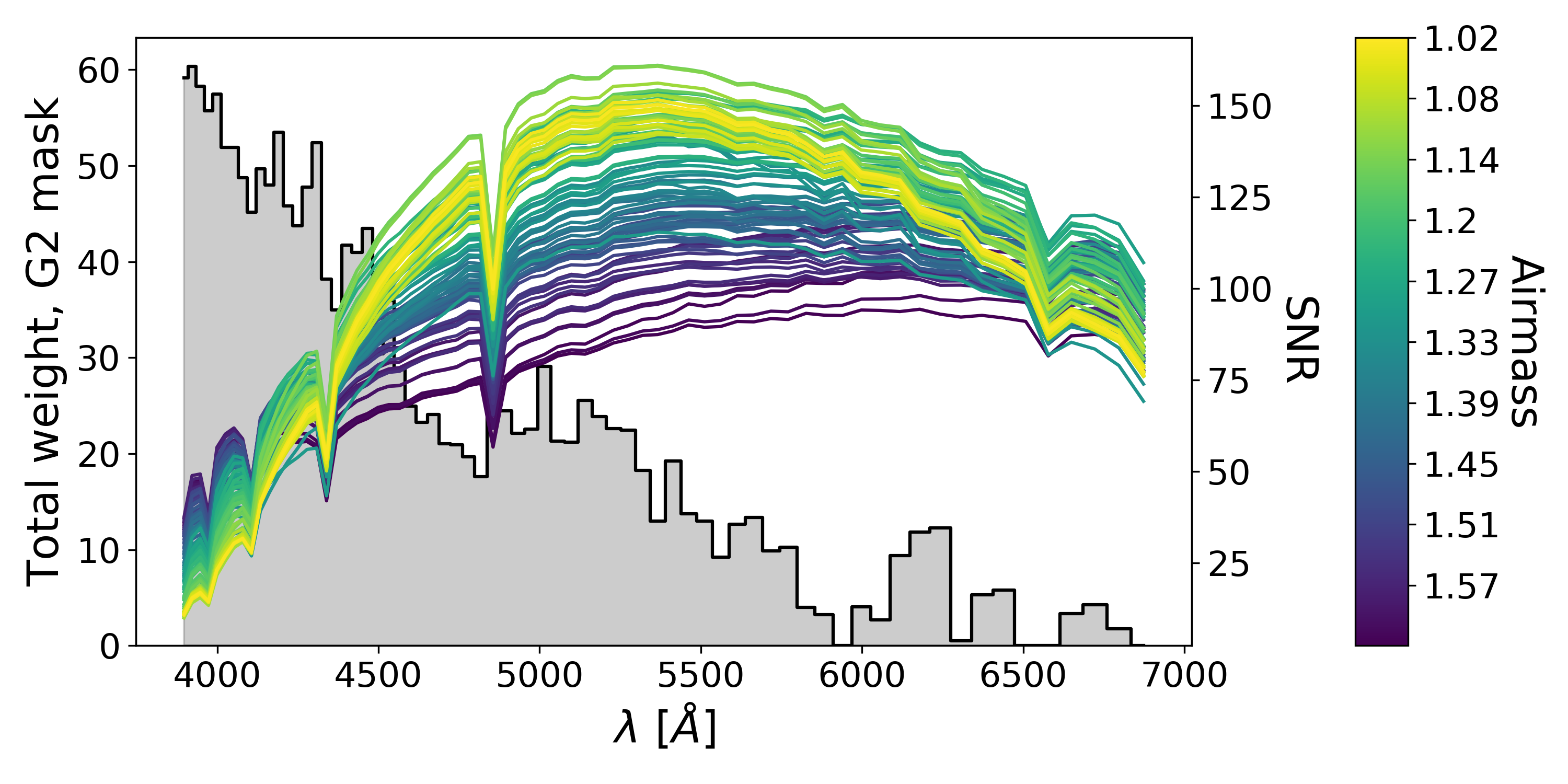}
\caption{\label{Fig:ADC}Signal-to-noise ratio as a function of airmass and wavelength (solid curves), and total weight within the mask for each spectral bin (gray histogram; accounting for the number of spectral lines and their depth only). Spectra acquired at lower airmass are expected to have higher signal-to-noise ratio throughout the spectrum, due to a lower optical depth of Earth atmosphere, but the malfunction of the ADC modifies this behaviour. This effect is particularly severe in the blue, where most of the information on the planet lies, as quantified by the weight in the binary mask. Indeed, while no change in observing conditions were noticeable in our run, the bluest orders of the lowest airmass spectra have a signal-to-noise ratio which is half compared to airmass 1.6, the maximum reached within our run.}
\end{figure*}

We observed the KELT-9 system in the framework of a Long-Term program (PI G. Micela) with HARPS-N and GIANO-B in GIARPS@TNG configuration \citep{Claudi2017}, as part of the GAPS project \citep{Covino2013}. For the present work, we only used the HARPS-N observations taken from the 22nd of July 2018 21:23 UT to the 23rd of July 2018 05:21 UT. The GIANO-B observations will be the target of a dedicated study. We collected 89 HARPS-N exposures, each with 180 seconds of integration. This is shorter compared to \cite{Hoeijmakers2018_k9} and \cite{Hoeijmakers2019}, who used an exposure time of 600 seconds. With this choice, the planet moved by at most $2.25~\mathrm{km~s^{-1}}$ during each exposure, which smeared the signal over 2.7 pixels. Considering the transit centred at phase 0, the planetary phases covered the range between 0.227 and 0.452, such that the planet was not occulted by the stellar disk.\\
We extracted and calibrated the spectra using the standard Data Reduction Software (DRS; version 3.7.1, \citealt{Dumusque2018}). To avoid the increase of correlated noise from data interpolation, we performed our analysis on the individual echelle orders (e2ds spectra), after correcting for the blaze function.  As previously reported by \cite{Borsa2019}, our observations were affected by a malfunction of the Atmospheric Dispersion Corrector that caused a deformation of the spectral energy distribution due to chromatic losses, and a concomitant loss of efficiency in the blue part of the spectra across the night (see Fig. \ref{Fig:ADC}). While we mitigated this effect with a custom color-correction, following \cite{Malavolta2017}, it is not possible to recover the lost signal-to-noise ratio at shorter wavelengths. We did not correct for telluric lines, because our analysis naturally excludes regions that are contaminated by them (Sec. \ref{subsec:CCF}). We then aligned the stellar spectra by removing the Barycentric Earth Radial Velocity motion, effectively shifting the spectra to the barycentric rest-frame of the solar system. This allowed us to build a high signal-to-noise ratio master stellar spectrum by: (1) rescaling every order to its average counts value and (2) computing a median in time for each order. The stellar motion induced by the planet amounts to about $0.2~\mathrm{km~s^{-1}}$ throughout the night, and does not significantly impact the shape of the stellar lines which are rotationally broadened by more than $100~\mathrm{km~s^{-1}}$. Since the planet moved in radial velocity by more than one pixel per exposure for most of the night, the resulting master spectrum contained the planetary lines only in minimal part. Each single e2ds spectrum was then divided by the master stellar spectrum, which removed the stellar lines. This procedure effectively provides the planet emission spectrum normalized to the stellar emission spectrum and planet continuum plus 1 (see Appendix \ref{sec:appendix_pipeline}). A high-pass filter was then applied to each of the resulting rows to remove residual discontinuities and low-order variations due to imprecise blaze or color correction (see Appendix \ref{sec:appendix_pipeline}). We found that the application of the high-pass filter enhanced the precision on the retrieved parameters by a factor of about 2. We finally applied a custom binary mask cross-correlation method (see Sec. \ref{subsec:CCF}).

\subsection{Line-weighted binary mask CCF}
\label{subsec:CCF}
\begin{figure*}
\centering
\includegraphics[width=17cm]{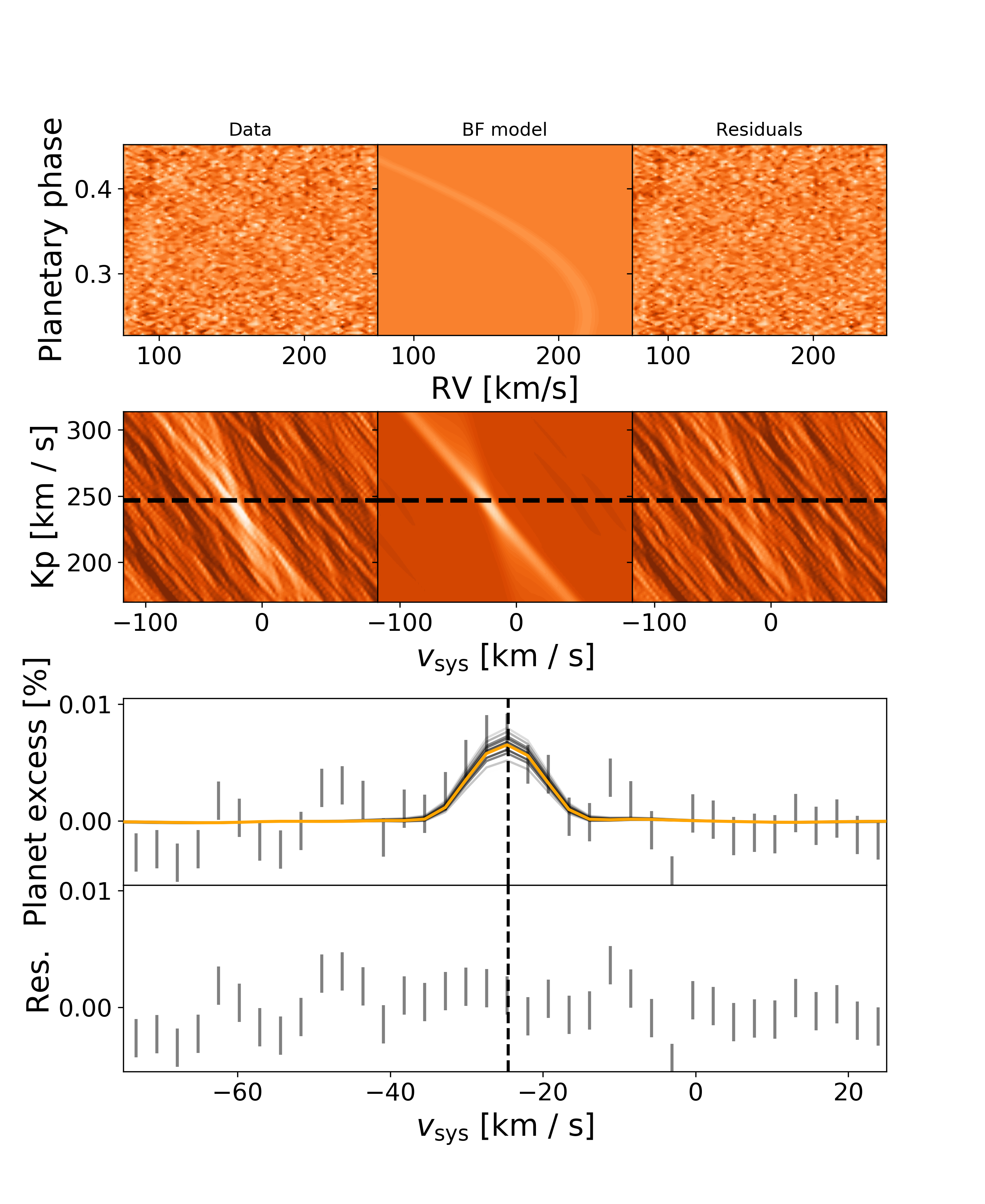}
\caption{ \label{Fig:Detection}Observed and modelled average planet emission line intersected by the G2 mask, and residuals between data and best fit model. \textit{Upper panel:} The exposure matrix in the region where we performed the fit. The curvature of the planetary trace is due to its overnight change in radial velocity compared to its host star. \textit{Middle panel:} $K_\mathrm{p}$--$v_\mathrm{sys}$ diagram for data, best fit model and residuals. The color scale is the same across the three panels, showing that the residuals map is clean in the region where the planet excess is localized. A horizontal, black, dashed line indicates the best fit value for $K_\mathrm{p}$. \textit{Lower panel:} The average data, model and residuals in the best fit planetary rest frame $K_\mathrm{p}$. Gray vertical lines are the data, with their uncertainties at 1 standard deviation, while the orange line is the model shown in the middle panel. Black lines are models deviating by less than $2\sigma$ from the best fit, while varying the iron abundance, with transparency proportional to their deviation. The bottom half of the panel shows the residualsfrom the best fit with the same y-axis. A black dashed vertical line shows the best fit systemic velocity. The average planetary line is in emission, and has a contrast of $84~\mathrm{ppm}$ compared to the continuum.}
\end{figure*}

With a temperature comparable to a K-dwarf, the atmosphere of KELT-9b should exhibit thousands of optical atomic transitions. The technique of cross-correlation \citep{Baranne1979, Sparks2002, Snellen2010} is best suited for their search  \citep{Hoeijmakers2019}. The technique was applied with different flavours (e.g. template matching, binary mask), and consists in stacking these thousands of planetary lines to abate the photon noise, which hinders their detection.\\
We adopted a CCF technique with a weighted binary mask \citep{Baranne1996, Pepe2002}\footnote{We do not normalize by the standard deviation. As such, our scheme is a cross-covariance in the statistical sense, but we call it cross-correlation following \cite{Baranne1996} and the signal processing nomenclature.}, where weights are attributed to individual spectral lines according to their information content (see Appendix \ref{sec:appendix_CCF}). It can employ the classic stellar binary masks used in the search of planets with the radial velocity method, as well as custom binary masks, and can be applied both to models and data. Compared to other cross-correlation schemes \citep{Snellen2010}, the binary mask approach preserves the contrast of the lines that it intercepts, which allows the comparison of the strength of different spectral features \citep{Pino2018b}. In practice, our technique extracts the average planet line normalized to the planet plus star continuum (which we call planet excess). This is similar to a least-squares deconvolution scheme (LSD, \citealt{Donati1997}), but without deconvolving the cross-correlation vector (with no loss of accuracy in the interpretation; Sec. \ref{sec:methods_models}). This average line profile can be used to directly extract observational properties of the planetary emission (Sec. \ref{sec:results}), but the extraction of parameters of the atmospheric structure requires the extra step of model comparison, for which we present a new method (Sec. \ref{sec:methods_models}).\\
Other works relied on similar definitions of the cross-correlation function \citep{Hoeijmakers2019}. However, they determined the weights on single pixels using model-injection, thus based on their information content, with the aim of reaching the highest signal-to-noise ratio on the planetary detection. In our approach, the binary mask attributes weights to single lines, as opposed to single pixels, with the advantage of reduced complexity and model-dependence. The consequently easier interpretation is obtained at the cost of a possible loss of signal-to-noise ratio, especially in the wings of the lines.\\
Since the planet has a temperature comparable to that of a star, in this work we adopted standard G2, K0 and K5 stellar masks provided by the DRS, optimized to extract radial velocities for planets orbiting stars for that spectral type. These masks are designed to exclude parts of the spectrum that are contaminated by telluric lines.\\ The results are mostly independent from the choice of the spectral type of the mask. This is likely because the masks share the strongest lines. Indeed, among the $1,000$ strongest lines in each mask, the majority of the lines are closer than $0.001~\mathrm{\AA}$, less than one tenth of a pixel. In percentage, the masks share $74.4\%$ (G2 vs K0), $82.6\%$  (G2 vs K5) and $84.4\%$ (K0 vs K5) of the strongest lines. In the following, we discuss the G2 mask case.\\
A CCF is computed for every exposure. The result is an `exposure matrix' which displays the planet trace in a diagram with radial velocity displacement from the stellar rest frame on the x-axis, and planetary phase (or exposure) on the y-axis (Fig. \ref{Fig:Detection}, upper panel). The fit (Sec. \ref{sec:methods_models}) was directly performed on this exposure matrix. However, we also display the results in the traditional $K_\mathrm{p}$--$v_\mathrm{sys}$ diagram, which visually highlights the presence or lack of a signal. In practice, we parametrized the planet orbit with a Keplerian velocity $K_\mathrm{p}$, appropriate for a circular orbit, and moved to the corresponding planet rest frame. Only the correct $K_\mathrm{p}$ aligns the individual CCFs, that are then summed. The maximum is thus found at the global radial velocity of the system (systemic velocity, $v_\mathrm{sys}$). This is conveniently represented in the $K_\mathrm{p}$--$v_\mathrm{sys}$ diagram (Fig. \ref{Fig:Detection}, middle panels).

\section{Observational results}
\label{sec:results}
In Fig. \ref{Fig:Detection}, we present the result of applying a G2 binary mask to the planet-to-star flux ratio. In practice, what we see is the average planet emission line intersected by the G2 binary mask normalized to the planetary and stellar continua. This emission is interpreted as due to the atmosphere of the planet.\\
The planetary atmospheric spectral feature, as seen through the G2 mask, has a contrast of $(84\pm1)~\mathrm{ppm}$ relative to the continuum. We obtained this by fitting a Gaussian curve to the planetary signal integrated over the exposures assuming the best fit $K_\mathrm{p}$ (see Fig. 2, lower panel; Sec. \ref{subsec:Best_fit}). The formal error is likely underestimated due to the presence of correlated noise. By replacing the formal error with the standard deviation far from the planet signal ($14~\mathrm{ppm}$, calculated at $-200~\mathrm{km~s^{-1}}<v_\mathrm{sys}<-100~\mathrm{km~s^{-1}}$), the signal-to-noise ratio of the detection is 6.\\
We then assumed that the continuum is the sum of the stellar and planetary continua (see Appendix \ref{sec:appendix_pipeline}). We further assumed that the stellar and planetary continua are blackbodies at temperatures of $10,000~\mathrm{K}$ and $4,570~\mathrm{K}$ \citep{Wong2019}, respectively. The contrast relative to the planetary continuum is then
\begin{equation}
\dfrac{F_{\mathrm{lines\,p}}}{F_{\mathrm{cont,\,p}}} = 84\cdot\left(1+\dfrac{F_{\mathrm{cont},\,\star}}{F_{\mathrm{cont\,p}}}\right)~\mathrm{ppm}\,,
\end{equation}
yielding $(40\pm5)\%$.\\
The planet excess appears in emission and not in absorption, which is an unambiguous sign of the presence of a thermal inversion in the atmosphere of the planet (see \citealt{Schwarz2015, Nugroho2017}; section \ref{sec:inversion}).

\section{Methods: extracting atmospheric parameters of KELT-9b}
\label{sec:methods_models}
The next step is extracting the planetary parameters from the cross-correlation function. This requires two ingredients: (1) a parametrized model for the exoplanet atmosphere (Sec. \ref{sec:model_grid}) and (2) a cross-correlation to likelihood mapping (Sec. \ref{sec:interpretation_scheme}). We also adapt the concept of contribution functions to the line-weighted binary mask CCF, to identify the pressure range probed by our analysis (Sec. \ref{sec:CCF_CF}).

\subsection{Model grid of KELT-9b atmosphere}
\label{sec:model_grid}
To compute planetary synthetic spectra, we developed a custom, line-by-line radiative transfer code that implements (1) opacities from the most important optical opacity sources in ultra hot Jupiters (\ion{Fe}{1}, \ion{Fe}{2}, \ion{Ti}{1}, \ion{Ti}{2}, $\mathrm{H^-}$; \citealt{Kitzmann2018, Arcangeli2018, Lothringer2019}), (2) equilibrium chemistry. LTE is assumed throughout the planetary atmosphere, and $\log g_\mathrm{p} = 3.3$. We neglected the reflected light component: from a theoretical standpoint, no reflective aerosols are expected in the atmosphere of the planet \citep{Kitzmann2018}; from an observational standpoint, due to the fast rotation of the host star \citep{Gaudi2017}, reflected spectral lines are broadened and thus difficult to detect with our continuum-normalized technique, that removes the majority of their signal. Furthermore, due to the polar orbit of the planet \citep{Gaudi2017}, the reflected stellar atomic lines would show a variable broadening approximately ranging between the intrinsic broadening of the stellar lines (quadrature) and the rotational broadening of the star (eclipse, 112 $\mathrm{km~s^{-1}}$), while the observed broadening is constant and consistent with the expected rotational broadening of the planet ($\sim6.63~\mathrm{km~s^{-1}}$). We further detail the radiative transfer code in Appendix \ref{sec:appendix_model}. Here we illustrate the parameter space explored.\\
Our synthetic spectra can be expressed as: 
\begin{equation}
S\left( \dfrac{\mathrm{VMR_{Fe}}}{\mathrm{VMR_{Fe}}_\star},\, \dfrac{\mathrm{VMR_{Ti}}} {\mathrm{VMR_{Ti}}_\star}, \,v_{\mathrm{sys}}, \,K_\mathrm{p} , \,v_\mathrm{rot,\,p},\, \mathrm{TP} \right),
\end{equation}
where $\mathrm{VMR_{Fe}} / \mathrm{VMR_{Fe}}_\star$ and $\mathrm{VMR_{Ti}} / \mathrm{VMR_{Ti}}_\star$ are the planetary to stellar iron and titanium volume mixing ratio, $v_\mathrm{rot,\,p}$ is the planetary rotational velocity assumed constant in the atmospheric region probed by the planetary emission lines, TP is a suitable parametrization of the temperature pressure profile. At our precision level, we expect retrieved abundances to be degenerate with the rotation rate (due to broadening) and the temperature profile, so that a full exploration of the parameter space is necessary to provide accurate constraints on each parameter. This is beyond the scope of this paper.\\
Instead, we focused on (1) determining which atomic species is mainly responsible for the observed average planetary emission line intersected by the G2 mask and (2) testing the hypothesis that the planet spectrum can be explained assuming abundances consistent with that of its host star. We could thus limit the parameter space by assuming that the planet is tidally locked ($v_\mathrm{rot,\,p}\cdot\sin i_p = 6.63~\mathrm{km~s^{-1}}$). Furthermore, we fixed the thermal profile to the self-consistent temperature profile  of KELT-9b that \cite{Lothringer2018} obtained by assuming a planetary metallicity equal to the stellar value and equilibrium chemistry. Under these reasonable assumptions, we produced three groups of models:
\begin{align}
\label{Eq:models}
\begin{split}
&S_\mathrm{Fe,\,Ti} \left( \dfrac{\mathrm{VMR_{Fe}}}{\mathrm{VMR_{Fe}}_\star},\, \dfrac{\mathrm{VMR_{Ti}}} {\mathrm{VMR_{Ti}}_\star},\, v_{\mathrm{sys}}, \,K_\mathrm{p}\right),\\
&\left.S_\mathrm{Fe} \left( \dfrac{\mathrm{VMR_{Fe}}}{\mathrm{VMR_{Fe}}_\star}, \,v_{\mathrm{sys}},\, K_\mathrm{p}\right)\right|_{\mathrm{VMR_{Ti}}/\mathrm{VMR_{Ti}}_\star = 0},\\
&\left.S_\mathrm{Ti} \left( \dfrac{\mathrm{VMR_{Ti}}}{\mathrm{VMR_{Ti}}_\star}, \,v_{\mathrm{sys}},\, K_\mathrm{p}\right)\right|_{\mathrm{VMR_{Fe}}/\mathrm{VMR_{Fe}}_\star = 0}.
\end{split}
\end{align}
$S_\mathrm{Fe}$ and $S_\mathrm{Ti}$ are obtained from $S_\mathrm{Fe,\,Ti}$ by removing titanium and iron, respectively. We fitted $v_{\mathrm{sys}}$ and $K_\mathrm{p})$, and  simultaneously varied $\mathrm{VMR_{Fe}} / \mathrm{VMR_{Fe}}_\star$ between $10^{-1}$ and $10^{3}$, and $\mathrm{VMR_{Ti}} / \mathrm{VMR_{Ti}}_\star$ between $3\cdot10^{-3}$ and $3\cdot10^{3}$. Since the host star KELT-9 has a metallicity between 0.7 and 2.7 times solar, higher volume mixing ratios seem unlikely. Lower volume mixing ratios would not be detectable at our precision level, and would thus not suffice to explain the data. 

\subsection{A new interpretation scheme for CCFs}
\label{sec:interpretation_scheme}
The strength of the cross-correlation signal depends on the quality of the match between the binary mask, or the model if used directly, to the data. For example, this can be quantified through peak signal to noise ratio. However, this approach is not statistically sound and can therefore not be used to estimate confidence intervals on planet parameters \citep{Brogi2016}. Alternatives exist, such as the Welch \textit{T}-test \citep{Brogi2013} or $\chi^2$--comparison based on model injection into data \citep{Brogi2016}, but they are computationally expensive. To overcome these challenges, \cite{Brogi2019} presented a cross-correlation to likelihood mapping to perform the comparison in a statistically sound framework (see also \citealt{Gandhi2019}), further generalized by \cite{Gibson2020}, while \cite{Fisher2019} proposed a different method based on a random forest approach.\\
Here, we propose a novel method to directly compare the cross-correlation of models and data. The procedure consists in simulating end-to-end synthetic observations, including the effects of data reduction. In the case of HARPS-N, this is facilitated by the small contamination from telluric lines. Furthermore, HARPS-N is a very stable instrument, built to acquire precise radial velocity observations. Consequently, our data reduction procedure is relatively simple. We are thus able to simulate end-to-end the effect of the data reduction process on synthetic e2ds HARPS-N generated from our models. This enables a direct comparison using a likelihood function, in a procedure similar to what \cite{Kochukhov2010} have previously suggested to interpret LSD profiles. We cross-checked our new method with the likelihood-mapping by \cite{Brogi2019}, finding good agreement (Appendix \ref{sec:appendix_test_retrieval}).\\
The first step is simulating the exposure matrix described in Sec. \ref{subsec:CCF} :
\begin{itemize}
\item We modelled the star using a PHOENIX model ($T_\mathrm{eff}=10,000~\mathrm{K}$, $\log g = 4.0$), and applied rotational broadening ($v_\mathrm{rot,\,\star}\cdot\sin i_\star = 111.8~\mathrm{km~s^{-1}}$, \citealt{Borsa2019}, linear limb darkening coefficient $\epsilon=0.6$).
\item We convolved each model emission spectrum of the exoplanet with a rotational kernel corresponding to the tidally locked solution ($v_\mathrm{rot,\,p}\cdot\sin i_p = 6.63~\mathrm{km~s^{-1}}$).
\item For each exposure $t_i$, we Doppler shifted every spectrum for a given orbital solution $(K_\mathrm{p}$, $v_\mathrm{sys})$. These velocities were parameters of the fit.
\end{itemize}
We then processed the simulated time-series through the procedure described in Appendix \ref{sec:appendix_pipeline}. The result was a model exposure matrix for each set of parameters ($K_\mathrm{p}$ and $v_\mathrm{sys}$; $\mathrm{VMR_\mathrm{Fe}}$ and  $\mathrm{VMR_\mathrm{Ti}}$), that we could directly fit to observations (see Fig. \ref{Fig:Detection}). Finally, we computed the Gaussian likelihood for radial velocities between $75~\mathrm{km~s^{-1}}$ and $252~\mathrm{km~s^{-1}}$, within which the planet trace is expected to be found, by\footnote{The method can be used with any other likelihood function}
\begin{equation}
\log \mathcal{L}=\sum_i \left[- \log \left( \sigma_i\sqrt{2\pi} \right) - \chi^2_i/2,\right],
\end{equation}
where $\sigma_i$ and $\chi^2_i$ are the error and $\chi^2$ associated to the data point $i$. We assumed that $\sigma_i$ is constant in radial velocity over an exposure, and measured it as the dispersion far from the expected position of the planet (radial velocities between $-200~\mathrm{km~s^{-1}}$ and $-100~\mathrm{km~s^{-1}}$).  The end result was a multi-dimensional $\log \mathcal{L}$ grid. We then employed different flavours of the likelihood test ratio to assess the significance of each model, to compare the models and to extract confidence intervals (see Appendix \ref{sec:appendix_basic_statistics} for practical details on how to do so).\\
This process is too slow to explore a large 4-dimensional grid of parameters. To speed it up, we found that: (1) rotational broadening can be included directly in the cross-correlated spectra; (2) instead of simulating all the exposures for each value of the couple ($K_\mathrm{p}$, $v_\mathrm{sys}$), the model exposure matrix can be directly shifted to simulate different values of the couple ($K_\mathrm{p}$, $v_\mathrm{sys}$) (see also \citealt{Brogi2019}). Practically, this assumes that the data reduction process effects on the planetary trace are independent of its $K_\mathrm{p}$ and $v_\mathrm{sys}$. We tested that both approximations do not cause a significant variation of the likelihood distributions.

\subsection{Contribution function of the cross-correlation function}
\label{sec:CCF_CF}
Since in our approach we are able to simulate the cross-correlation function of each model, for a given assumed atmospheric structure it is possible to directly study the location in pressure where the cross-correlation signal originates from. This can be described with a `contribution function to the cross-correlation function at the surface'. To our knowledge, this is the first time that the contribution function is adapted to the context of high spectral resolution observations of planetary atmospheres. We define it here by analogy with the classic contribution function to the flux at the surface.\\
Following e.g. \cite{Irwin2009} and \cite{Malik2019} we define the contribution functions as the contribution of each discrete layer in our model to the flux at the surface of the planetary atmosphere. In our case, we locate the `surface' high-up in the optically thin region of the planet atmosphere, from which the photons escape and reach the observer. If every layer $n$ emits an intensity $\Delta_n \mathrm{I}(\mu)$ in a direction $\mu=\cos \theta$, we can write:
\begin{equation}
\label{Eq:CF_intensity}
\mathrm{I}(\mu) = \sum_n\left[\Delta_n \mathrm{I}(\mu)\exp\left(-\tau_n/\mu\right)\right]\,,
\end{equation}
where $\tau_n$ represents the optical depth above layer $n$, and $\Delta_n\mathrm{I}$ is computed according to the linear in optical depth approximation \citep{Toon1989}. The $n$-th term in square brackets on the right-hand side of the equation is the contribution function of layer $n$.\\
We now apply the cross correlation at the left hand and right-hand side of Eq. \ref{Eq:CF_intensity}. The sum over $n$ atmospheric layers can be commuted with the sums contained in our definition of CCF (Eq. \ref{Eq:CCF_discrete}). As a result, we can write:
\begin{equation}
\label{Eq:CF_CCF}
\begin{split}
&\mathrm{CCF}(\mathrm{\mathrm{I(\mu)}}) = \mathrm{CCF}\left(\sum_n\left[\Delta_n \mathrm{I}(\mu)\exp\left(-\tau_n/\mu\right)\right]\right) =\\
&= \sum_n \mathrm{CCF}\left(\left[\Delta_n \mathrm{I}(\mu)\exp\left(-\tau_n/\mu\right)\right]\right)\,.
\end{split}
\end{equation}
By extension, the $n$ terms in square brackets in the right-hand side of Eq. \ref{Eq:CF_CCF} represent the ``contribution functions of the cross-correlation function'' of each layer. Physically, they represent the contribution to the intensity as a function of radial velocity rather then wavelength from every atmospheric layer.\\
With this definition, it is trivial to identify the pressure range that can be probed with a line-weighted binary mask CCF of high spectral resolution observations. Furthermore, for a given model, the contribution functions of the CCF inform us on which pressure layers more tightly constrain the radial velocity of the planet. By integrating over $\mu$ one obtains expressions for the flux.

%
%
%
%
%%%%%%%%%%%%%%%%%%%%%%%%%%%
%%%%%% INTERPRETATION %%%%%%%%%%%
%%%%%%%%%%%%%%%%%%%%%%%%%%%
%
%
% 

\section{Results from model comparison}
\label{sec:results_models}
In the following, we provide our interpretation of the average planet line intersected by the G2 mask based on model comparison.

\subsection{Fit with line weighted binary mask}
\label{subsec:Best_fit}
\begin{figure*}
\centering
\includegraphics[width=17cm]{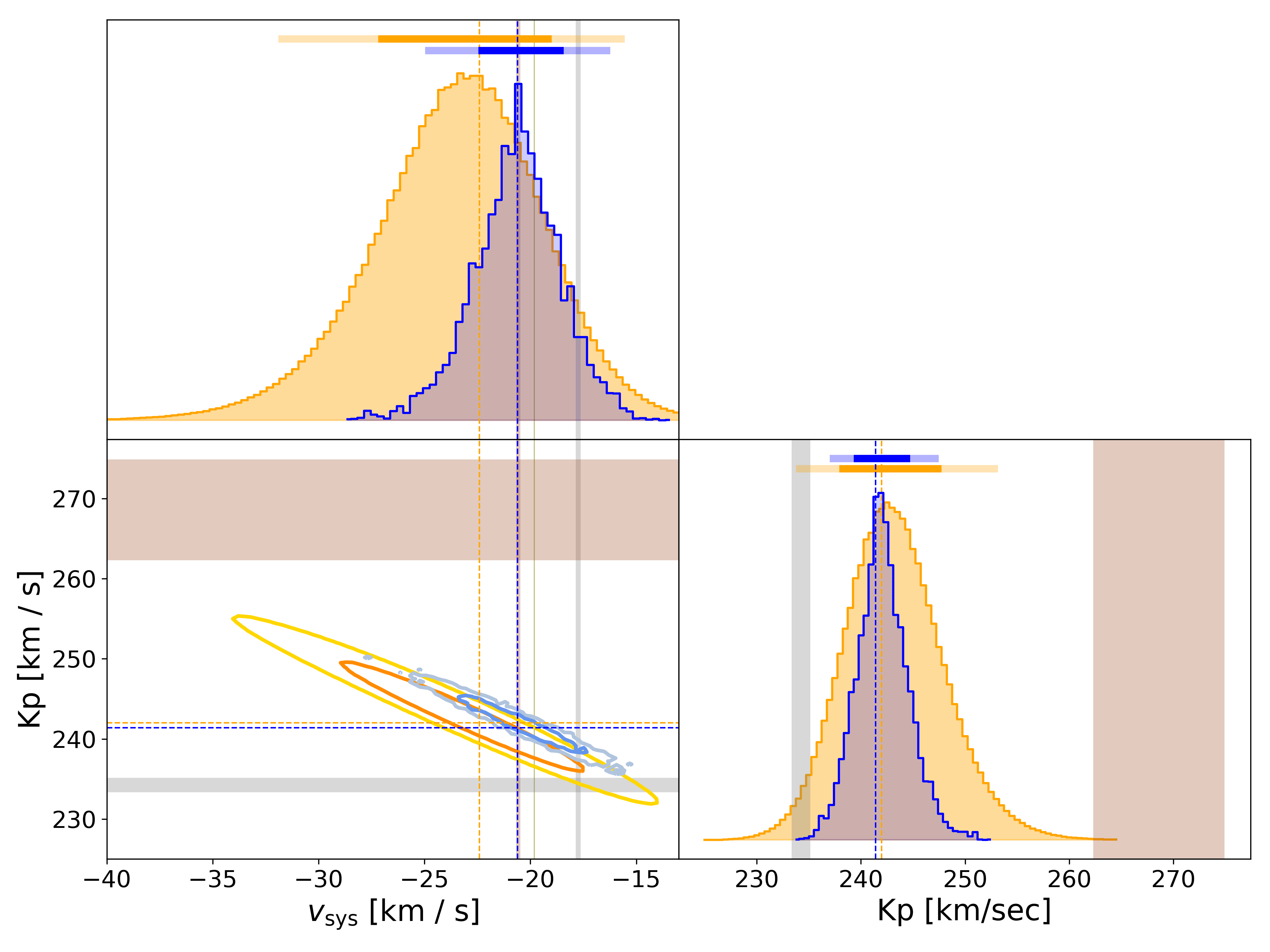}
\caption{Marginalized likelihood distributions for $v_\mathrm{sys}$ and $K_\mathrm{p}$ for the line weighted binary mask (orange) and for the \cite{Brogi2019} approach (blue). Dark and light orange (blue) horizontal bars denote the $1\sigma$ and $2\sigma$ confidence levels. Orange (blue) dashed lines indicate the best-fit value. Shaded areas denote the literature values by \cite{Yan2018} (sienna), \cite{Hoeijmakers2019} (gray) and \citep{Borsa2019} (olive). \cite{Borsa2019} only measure $v_\mathrm{sys}$. Our distributions for $v_\mathrm{sys}$ is consistent with the literature, while we deviate from the $K_\mathrm{p}$ value by \cite{Yan2018} by about $3\sigma$. \label{Fig:likelihood_velocities}}
\end{figure*}

We first identified which among the models defined in Eq. \ref{Eq:models} best explains the data. The model containing only lines from neutral and ionized titanium and no atmospheric iron, $S_\mathrm{Ti}$, has maximum likelihood at the highest allowed abundances of titanium. This suggests that titanium lines are too weak to explain the observed emission lines even when $\mathrm{VMR_{Ti}}=3,000\cdot\mathrm{VMR_{Ti_\star}}$. We then compared $S_\mathrm{Ti}$ to the full model  $S_\mathrm{Fe,\,Ti}$ with a likelihood test ratio (see Appendix \ref{sec:appendix_basic_statistics}), and found that it can be excluded at $4.3\sigma$. When limiting the maximum abundance of titanium to 100 times the stellar value, the model can be excluded at $5.1\sigma$. As a further indication that iron is necessary to explain the observed emission line, we calculated the difference in Bayesian Information Criterion (BIC, \citealt{Liddle2007}) and found that $\min \left[\mathrm{BIC} (S_\mathrm{Ti})\right]$ = $\min \left[\mathrm{BIC} (S_\mathrm{Fe,\,Ti})\right]+10$. The difference increases to 17.5 when limiting the maximum abundance of titanium to 100 times the stellar value, indicating strong preference for the presence of iron.\\
In a similar fashion, we compared the model containing only lines from neutral and ionized iron and no atmospheric titanium, $S_\mathrm{Fe}$, to the full model. In this case, the null hypothesis that $S_\mathrm{Fe}$ is the true model can not be excluded ($<0.1\sigma$). Furthermore, it is strongly favoured by the BIC test, with $\min \left[\mathrm{BIC} (S_\mathrm{Fe})\right]$ = $\min \left[\mathrm{BIC} (S_\mathrm{Fe,\,Ti})\right]-8.7$, which penalizes the presence of an additional parameter in $S_\mathrm{Fe,\,Ti}$. We thus adopted $S_\mathrm{Fe}$ as our nominal model to derive planetary parameters (see Table \ref{Tab:model_comparison_iron_and_titanium}).\\
The best fit parameters are $K\mathrm{p}=242~\mathrm{km~s^{-1}}$, $v_\mathrm{sys}=-22.5~\mathrm{km~s^{-1}}$, $\mathrm{VMR_{Fe}}=30\cdot\mathrm{VMR_{Fe_\star}}$. The model is a very good match to the data, as evidenced by a reduced $\chi^2 =6128/5874=1.043$ and by residuals within the statistical fluctuations (Fig. \ref{Fig:Detection}). We computed the significance of the model by performing a likelihood test ratio, comparing it to the case of null detection $\mathrm{VMR_{Fe}}=0$ (a straight line; see Appendix \ref{sec:appendix_basic_statistics}). The  detection is significant at $6.15\sigma$. As a further test, we computed that the BIC value of our best fit model is lower by 24.5 compared to the null detection, indicating a strong preference for the presence of iron. The $1\sigma$ confidence intervals for the three parameters (see Appendix \ref{sec:appendix_basic_statistics}) are $238~\mathrm{km~s^{-1}}< K_\mathrm{p} < 247.5~\mathrm{km~s^{-1}}$, $-32 < v_\mathrm{sys} < -19$ and $10< \mathrm{VMR_{Fe}} / \mathrm{VMR_{Fe}}_\star < 150$ (compatible with a few times the stellar value at $2\sigma$). \\
Finally, we compared our nominal model $S_{\mathrm{Fe}}$ with two models where we suppressed lines by neutral and ionized iron respectively. These two models are not formally nested in $S_{\mathrm{Fe}}$, and we compared instead the significance yielded by the best fit parameters for each model. When only neutral iron is present, the results are nearly indistinguishable from the full model $S_{\mathrm{Fe}}$, with a similar significance, best fit and confidence interval. On the other hand, when only ionized iron is present, the best fit is found at the upper limit of $\mathrm{VMR_{Fe}}=1,000\cdot\mathrm{VMR_{Fe_\star}}$ and has a much lower significance of $3.1\sigma$. In this case, the BIC test favours the null detection, indicating that the ionized iron lines intersected by the G2 mask are too weak to explain the observed planetary feature alone (see Table \ref{Tab:model_comparison_iron_only}).\\
We also applied the method by \cite{Brogi2019} to perform an independent test (see Appendix \ref{sec:appendix_test_retrieval}). In this case, we fixed the abundance to its best-fit value, and retrieved $K_\mathrm{p}$ and $v_\mathrm{sys}$ and a scale factor which is a proxy for abundance. The results are in good agreement with those found with our novel framework (see Fig. \ref{Fig:likelihood_velocities}, and Appendix \ref{sec:appendix_test_retrieval}).

\begin{table}
\label{Tab:model_comparison_iron_and_titanium}
\begin{center}
\caption{Comparison of models containing iron or titanium lines.}
\begin{tabular}{ccccc}
\hline 
\hline 
 & $\Delta \mathrm{BIC}$ with & LRT with  \\ 
 &  $S_\mathrm{Fe,\,Ti}$  & $S_\mathrm{Fe,\,Ti}$  \\ 
\hline	
$S_\mathrm{Fe}$ & -8.7 & $<0.1\sigma$\\ 
$S_\mathrm{Ti}$  & +10 & $4.3\sigma$\\
\hline 
\end{tabular} 
\end{center}
\textbf{Notes.} The LRT metric indicates that a model containing neutral and ionized iron ($S_\mathrm{Fe}$) explains the data as well as a model containing also neutral and ionized titanium. On the other hand, a model containing only lines from neutral and ionized titanium ($S_\mathrm{Ti}$) does significantly worse. Furthermore, the BIC difference favours the model containing only neutral and ionized iron, and no titanium, due to the smaller number of free parameters. We thus adopt $S_\mathrm{Fe}$ as fiducial model.
\end{table}

\begin{table}
\label{Tab:model_comparison_iron_only}
\begin{center}
\caption{Comparison of models containing neutral iron lines, ionized iron lines or both.}
\begin{tabular}{ccccc}
\hline 
\hline 
 & $\Delta \mathrm{BIC}$ with & LRT with \\ 
 &  null detection  & null detection \\ 
\hline	
Neutral and ionized  & -25 & $6.15\sigma$\\ 
iron ($S_\mathrm{Fe}$) & &\\
Neutral iron only & -25 & $6.15\sigma$\\ 
Ionized iron only  & +5 & $3.1\sigma$\\
\hline 
\end{tabular} 
\end{center}
\textbf{Notes.} The LRT metric indicates that a model containing only ionized iron has a lower significance compared to the null detection. Although the significance is still at the $3\sigma$ level, this occurs at the upper limit of the allowed iron abundances (1,000 times solar), and the BIC test significantly disfavours this model compared to a flat line. Neutral iron is thus necessary to explain the data under our assumptions. Furthermore, The addition of ionized iron does not significantly improve the fit, or significantly change the inferred iron abundance.
\end{table}

\subsection{Reference frame of the signal}
The comparison at face-value of the joint probability distributions and the marginalized 1D probabilities reveals that our results are consistent with all literature values of the systemic velocity (\citealt{Gaudi2017} adopted by \citealt{Yan2018}, \citealt{Hoeijmakers2019} and \citealt{Borsa2019}; see Table \ref{Tab:literature_vsys_Kp}). While these authors reported  individual precisions around $0.1~\mathrm{km~s^{-1}}$, the measured values are significantly discrepant, spanning a range of about $3~\mathrm{km~s^{-1}}$. Further analysis is required to pinpoint the origin of this discrepancy. We thus attributed an error of $3~\mathrm{km~s^{-1}}$ to the single measurements to account for an unknown systematic effect.  With this assumption, the average $v_\mathrm{sys,\,\star} = -19 \pm 3~\mathrm{km~s^{-1}}$ is compatible within one sigma with our result ($\Delta_{v_\mathrm{sys,\,\star}} = 3.5^{+5.5}_{-4.5}~\mathrm{km~s^{-1}}$ and $\Delta_{v_\mathrm{sys,\,\star}} = 1^{+3}_{-4}~\mathrm{km~s^{-1}}$ for the line weighted binary mask and the \citealt{Brogi2019} approaches respectively).\\
Furthermore, deviations between $K_\mathrm{p}$ measured from atomic metal lines in emission (our work) and in transmission \citep{Hoeijmakers2019} are in agreement at the $2\sigma$ level. However, the $K_\mathrm{p}$ value measured by \cite{Yan2018} on the $\mathrm{H}\alpha$ line is in tension with the $K_\mathrm{p}$ measured on the metal lines ($\Delta K_\mathrm{p} = 27^{+7.5}_{-8}~\mathrm{km~s^{-1}}$ and $\Delta K_\mathrm{p} = 27.5\pm 6~\mathrm{km~s^{-1}}$ for the line weighted binary mask and \citealt{Brogi2019} approaches respectively). We explored the possibility that this difference is of astrophysical origin, due to the fact that the hydrogen and iron lines probe different regions of the atmosphere. \cite{Yan2018} report that the $\mathrm{H}\alpha$ line approaches but does not reach the Roche lobe. Furthermore, the $\mathrm{H}\alpha$ line has a symmetrical profile. Therefore, it is likely generated below the exosphere, in the part of the atmosphere gravitationally bound to KELT-9b. Any relative motion between the gas components probed by observations should thus be subsonic. By assuming the adiabatic coefficient of a monoatomic gas, the temperature profile by \cite{Lothringer2018} and the mean molecular weight from our model, we obtain that the sound speed ranges between $6.5~\mathrm{km~s^{-1}}$ and $8.5~\mathrm{km~s^{-1}}$. If it was of astrophysical origin, the difference between the semi-amplitude measured by \cite{Yan2018} and our measurement would thus be larger then the sound speed (although only marginally in the case of the line weighted binary mask), which is unlikely. Further dedicated work is necessary to consistently explain these observations.

\begin{table}
\label{Tab:literature_vsys_Kp}
\begin{center}
\caption{Literature and derived $v_\mathrm{sys}$ and $K_\mathrm{p}$ values.}
\begin{tabular}{ccccc}
\hline 
\hline 
 & $v_\mathrm{sys}~[km~s^{-1}]$  & $K_\mathrm{p}~[km~s^{-1}]$ \\ 
\hline	
\cite{Yan2018} & $-20.6\pm0.1^\mathrm{a}$ & $269^{+6.5}_{-6}$ &\\ 
\cite{Borsa2019}  & $-19.81\pm0.02$ & --\\ 
\cite{Hoeijmakers2019} & $-17.7\pm0.1$ & $234.24\pm0.9$ & \\ 
This work, G2 mask & $-22.5^{+3.5}_{-4.5}$ &$242^{+5}_{-4}$\\
This work, & $-20.5^{+2}_{-1.5}$ & $241.5^{+3}_{-2}$\\
\cite{Brogi2019} technique  & &\\
\hline 
\end{tabular} 
\end{center}
\textbf{Notes.} The error bars indicate $1\sigma$ intervals reported in the literature, or on the 1D marginalized likelihoods. Our results are broadly consistent with the literature, with the exception of  $K_\mathrm{p}$ measured by \cite{Yan2018}. When both are measured from the planetary spectrum, systemic velocity and Keplerian velocity are correlated, as evident from Fig. \ref{Fig:likelihood_velocities}, where the 2D confidence intervals are reported.\\
$^\mathrm{a}$ Taken from \cite{Gaudi2017}.

\end{table}

\begin{figure*}
\centering
\includegraphics[width=17cm]{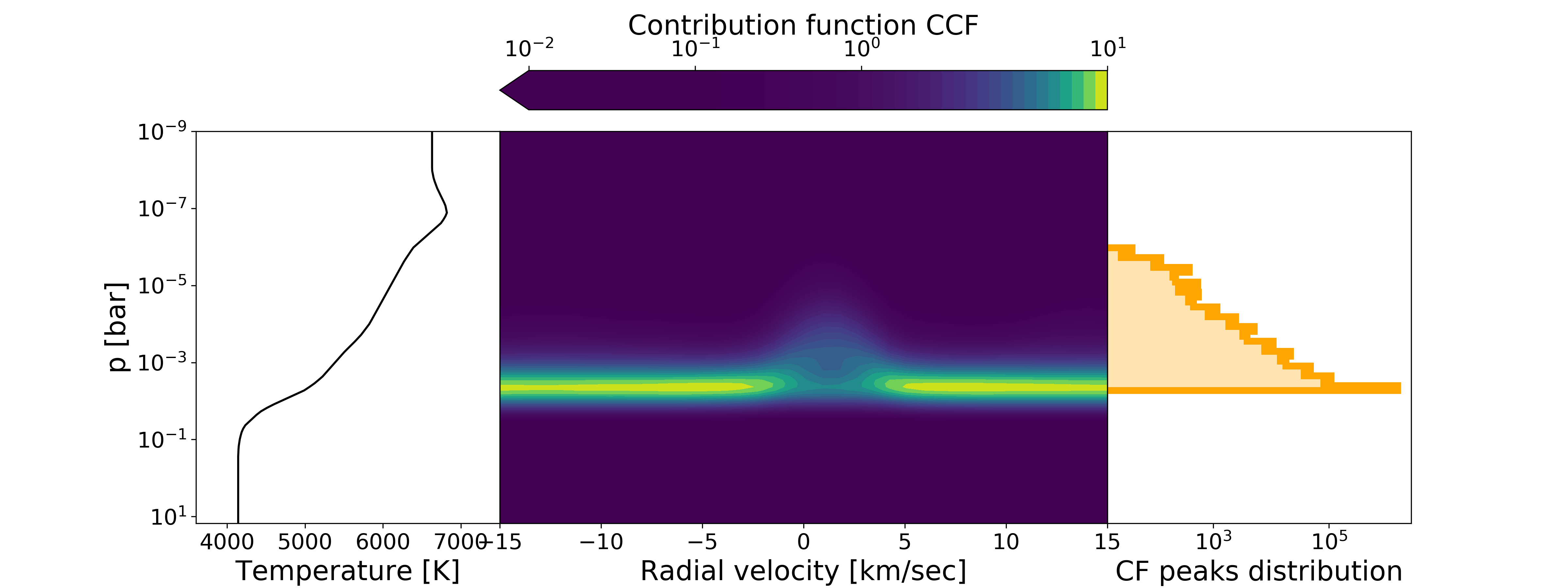}
\caption{ \label{Fig:CF} Diagnostics of the contribution functions of a model assuming stellar iron abundance and the temperature-pressure profile by \cite{Lothringer2018} (shown in the \textit{left panel}). \textit{Central panel:} Cross-correlation of the contribution function, performed layer by layer. This indicates the relative contribution to the flux as a function of radial velocity rather than wavelength (see Appendix \ref{sec:CCF_CF}). The continuum of the cross-correlation function is located around a few bars, and is due to absorption by $\mathrm{H}^{-}$. The peak of the CCF is mostly sensitive to pressure levels around $10^{-3/-5}~\mathrm{bar}$. \textit{Right panel:} For every wavelength channel in our model, we look for the location in pressure of the maximum of the contribution function, and produce an histogram. The two separated peaks show that the continuum originates at the pressure of a few bars, and that the core of most of the iron lines are originated between pressures of $10^{-3/-5}~\mathrm{bar}$ (mind the logarithmic scale of the counts).}
\end{figure*}

\section{Discussion}

\label{sec:discussion}

\subsection{A temperature inversion in the day-side of KELT-9b}
\label{sec:inversion}
The average planet line intersected by the G2 mask is in emission, which can only be explained if a thermal inversion is present in the atmosphere of KELT-9b. This conclusion is model-independent, since it only hinges on the sign of the planetary lines, which is preserved by our analysis.\\
We calculated the contribution functions to the CCF of the model adopting the thermal inversion by \cite{Lothringer2018} and solar iron abundance (Sec. \ref{sec:CCF_CF}). The emission from the neutral iron line cores originates between $10^{-3}~\mathrm{bar}$ and $10^{-5}~\mathrm{bar}$ (see Fig. \ref{Fig:CF}). This is higher-up compared to the $\sim30~\mathrm{mbar}$ region probed by \cite{Hooton2018}, who reported an evidence of inversion using ground-based photometry. It is also well within the inverted region of the atmosphere, found above the region of absorption of stellar irradiation and located between 1 and $100~\mathrm{mbar}$ in the optical region probed by HARPS-N \citep{Lothringer2018}.\\
For hot Jupiters with equilibrium temperature larger than $1,600~\mathrm{K}$, molecules with strong optical opacities such as TiO and VO are expected to be in the gas phase causing a temperature inversion below $0.1~\mathrm{bar}$ \citep{Hubeny2003, Fortney2008}. For the higher temperatures experienced by ultra-hot Jupiters, most molecules are dissociated, so these species become irrelevant for the thermal inversion. Instead, atomic metals and metal hydrides are capable of absorbing enough short wavelength irradiation to heat up the atmosphere. In particular, the bound-bound transitions of neutral iron absorbs significantly long-ward of $0.3~\mathrm{\mu m}$, and the bound-free transitions absorbs the high-energy flux short-ward of $0.3~\mathrm{\mu m}$ \citep{Sharp2007}. This is enough to create a thermal inversion at $10~\mathrm{mbar}$ \citep{Lothringer2018}. Higher up, around $0.5~\mathrm{mbar}$, iron is mostly found in its ionized form due to the high atmospheric temperature.\\
A second important factor that contributes to the formation of thermal inversions is the lack of molecules with near-infrared opacities, able to radiatively cool the atmosphere. This can be caused by high C/O atmospheres \citep{Molliere2015, Gandhi2019} and/or by thermal dissociation \citep{Lothringer2018, Parmentier2018, Arcangeli2018}, with the latter scenario predicted to be important in ultra-hot Jupiters \citep{Lothringer2019, Malik2019}.

\subsection{On the chemical composition of KELT-9b}
Ultimately, we conclude that the average KELT-9b emission line intersected by the G2 mask can be explained with neutral iron as predicted by equilibrium chemistry, with iron abundance compatible with a few times that of the host star. However, our results do not imply a lack of ionized iron lines or other species. Furthermore, with the current analysis, our confidence intervals on $\mathrm{VMR_{Fe}}$ are likely too narrow. This is because (1) we fixed the thermal profile and rotation rate and (2) the choice of a specific mask inherently biases the results by selecting specific pixels within the spectrum.\\
Looking forward, an application of our method with additional line weighted masks sensitive to different lines, and with additional models exploring different thermal profiles, may provide an avenue to measure an [Fe/H] potentially representative of the whole planetary atmosphere. Indeed, iron condenses through a simple phase transition, passing to the liquid or solid state. When present, iron clouds effectively remove most of the iron from the atmosphere above them \citep{Visscher2010}. The mere presence of iron lines in the atmosphere of a planet indicates the likely absence of deep iron clouds, suggesting that the measured abundance may be representative of the global iron abundance in the planetary atmosphere.

\subsection{Comparison of transmission and emission spectroscopy of iron lines}
The transmission spectrum of the planet atmosphere probes its terminator region, where lower temperatures are expected, which could reflect in different chemical properties of the atmosphere. \cite{Hoeijmakers2019} reported absorption from neutral iron at the terminator of KELT-9b at the millibar level by assuming the pressure level of the planetary continuum. This would be at a similar pressure compared to what we report here looking at the day-side emission line. \cite{Hoeijmakers2019} also reported the detection of ionized iron lines, which they estimated to be at the $\mu\mathrm{bar}$ level, higher up compared to the pressure level where neutral iron emission lines originate from in our scenario.  The combination of these results covers three orders of magnitude in pressure, although we highlight that we find no evidence for ionized iron with our analysis.\\
From a geometrical standpoint, transmission spectroscopy is sensitive to lower densities compared to emission spectroscopy.  Therefore, the combination of the transmission and emission findings could suggest that neutral iron is depleted at around 0.1 millibar at the terminator compared to the day-side atmosphere of the planet. However, we emphasize that for both the emission and transmission studies, the pressure levels where spectral features originate were calculated by making assumptions regarding the temperature profile and gravity of the planet, and assuming a hydrostatic profile for the atmosphere. Further work to explore the effect of these assumptions is required to properly combine the data sets. Nevertheless, this comparison demonstrates the potential to characterize the 3D structure of the atmosphere of exoplanets by studying them at high spectral resolution both in transmission and emission.

%
%
%
%
%%%%%%%%%%%%%%%%%%%%%%
%%%%%% APPENDIX %%%%%%%%%%
%%%%%%%%%%%%%%%%%%%%%%
%
%
%

\appendix

\section{A physical interpretation of the cross-correlation function}
\label{sec:appendix_pipeline}
In this Appendix we aim to provide a physical understanding of the planetary excess observed in Fig. \ref{Fig:Detection}. This step is fundamental to properly set-up the simulations to be compared with data. We thus describe in mathematical detail (1) how the observations are related to the planetary and stellar spectrum and (2) the steps undertaken to normalize the spectral observations described in \ref{sec:methods:observations_and_data_reduction}. The steps involved:
\begin{enumerate}
\item a color-correction, to mitigate chromatic losses that change the spectral shape observed over the night. This was particularly important for our observations, due to the failure of the ADC which corrects part of these effects at the telescope level;
\item a rescaling of the spectrum to its continuum in every order, to account for variations of the signal-to-noise overnight;
\item a normalization to the stellar spectrum, obtained directly from the data, to remove stellar lines.
\end{enumerate}

\subsection{Relation between observations and planetary and stellar spectra}
HARPS-N records combined-light observations of the star and planet system at any given time in units of photoelectron counts $\mathrm{C}(\lambda_n, t_i)$, split in orders $m$ (e2ds spectra). In other words, the information they contain is the total energy deposited in each pixel $n$ during the exposure $i$. On the other hand, both the PHOENIX models and our radiative transfer code output a spectral flux density, i.e. energy per unit wavelength per unit area per unit time $\mathrm{F}(\lambda_n, t_i)$. We assume that these quantities are related by:
\begin{equation}
\label{Eq:counts_flux}
\mathrm{C}(\lambda_n, t_i) = \mathrm{LSF} * \left\{ \left[R_p^2 \cdot \mathrm{F_\mathrm{p}}(\lambda_n, t_i) + R_\star^2 \cdot \mathrm{F_\mathrm{\star}}(\lambda_n)\right]\cdot \mathcal{A}(\lambda_n, t_i) \cdot \mathcal{B}(t_i) \right\}(\lambda_n, t_i)  \cdot \Delta t_i \cdot \frac{A_\mathrm{tel}}{d^2} \cdot \Delta \lambda_n \cdot \mathcal{G}\,,
\end{equation}
where $\Delta t_i$ is the exposure time, $A_\mathrm{tel}/d^2$ is the ratio of the area of the telescope to the distance of the system squared, $\Delta \lambda_n$ is the wavelength range  covered by the pixel $n$, and $\mathcal{G}$ is a gain factor. We added two factors $\mathcal{A}$ and $\mathcal{B}$ to indicate chromatic losses ($\mathcal{A}$; e.g. chromatic losses at the fibre entrance due to atmospheric dispersion) and overall flux losses ($\mathcal{B}$; e.g. seeing variations, airmass effects). While $\mathcal{B}$ is a simple scaling factor between the exposures, $\mathcal{A}$ changes the shape of the spectrum in each exposure.\\
The relation is non-linear because the Line Spread Function of the spectrograph is convolved with the received spectral flux density, and the planet and star fluxes are already convolved with the respective rotational broadening kernel. In the rest of the discussion we assume that $\mathrm{C}(\lambda_n, t_i)$ is proportional to $\mathrm{F}(\lambda_n, t_i) =R_p^2 \cdot \mathrm{F_\mathrm{p}}(\lambda_n, t_i) + R_\star^2 \cdot \mathrm{F_\mathrm{\star}}(\lambda_n)$, which we find true at a precision better than 0.1 parts-per-million (see also \citealt{Pino2018a}).\\
After having related observations and models, we turn to understanding how the data reduction process that we follow impacts the models in mathematical detail. With this next passage, we get a physical understanding of what the observed cross-correlation function (Fig. \ref{Fig:Detection}) means.

\subsection{Preparation of spectra for cross-correlation}
\label{sec:appendix_prepararion_of_spectra}
To combine the spectra in order to increase the signal-to-noise ratio, and properly extract the planet signal, the data reduction process aims at removing the time and wavelength dependence of $\mathcal{A}(\lambda_n, t_i)$ and $\mathcal{B}(t_i)$.\\
The first step is color-correction, which removes the wavelength dependence of these multiplicative noise factors. Colour-correction is performed relative to a template, for which we used the first spectrum of the night, where the ADC was performing the best. We produced a low-resolution (LR) version of each spectrum, with one single point in every order. To remove temporal variations, every low-resolution spectrum is rescaled to its spectral order 48 ($5,580~\mathrm{\AA} < \lambda < 5,640~\mathrm{\AA}$):
\begin{equation}
\label{Eq:low_resolution_spectrum}
\mathrm{C_{LR}}(\lambda_n, t_i) = \dfrac{\langle \mathrm{C(\lambda_n, t_i)}\rangle_{\mathrm{order\,}m}}{\langle \mathrm{C(\lambda_n, t_i)}\rangle_{\mathrm{order\,}48}} = \dfrac{\langle \mathrm{F}(\lambda_n, t_i)\mathcal{A}(\lambda_n, t_i)\rangle_{\mathrm{order\,}m}}{\langle \mathrm{F}(\lambda_n, t_i)\mathcal{A}(\lambda_n, t_i)\rangle_{\mathrm{order\,}48}}\,
\end{equation}
where we used Eq. \ref{Eq:counts_flux}, angular brackets indicate average between pixels 1024 and 3072 of each order and we simplified several wavelength independent factors. By assuming that the factor $\mathcal{A}(\lambda_n, t_i)$ is approximately a constant $\mathcal{A}(t_i)_m$ over an order $m$ and that the planet flux is small compared to the star, we obtain a residual curve for each exposure $i$:
\begin{equation}
\label{Eq:residuals_color_correction}
\dfrac{\mathrm{C_{LR}}(\lambda_n, t_i)}{\mathrm{C_{LR,\,templ}}(\lambda_n)} = \dfrac{\mathcal{A}(t_i)_m}{\mathcal{A}_{\mathrm{templ},\,m}}\,.
\end{equation}
Eq. \ref{Eq:residuals_color_correction} represents the variation of each spectrum compared to a template only due to the color effect, and needs to be removed from the spectra. We determined that an interpolation with a sixth order spline in wavelength at each $\lambda_n$ for each exposure minimizes the residuals. We then obtain a color corrected version of $\mathrm{C}$ by dividing Eq. \ref{Eq:counts_flux} by Eq. \ref{Eq:residuals_color_correction}:
\begin{equation}
\mathrm{C_{cc}(\lambda_n, t_i)} = \mathrm{LSF} * \left\{\mathrm{F}(\lambda_n, t_i)\cdot \mathcal{A}_\mathrm{templ}(\lambda_n)\cdot \mathcal{B}(t_i) \right\} \cdot \Delta t_i \cdot \frac{A_\mathrm{tel}}{d^2} \cdot \Delta \lambda_n \cdot \mathcal{G} \,.
\end{equation}
Some extra time-dependent (and wavelength independent) factors have been absorbed in $\mathcal{B}(t_i)$. We stress that color correction only ensures that the relative shape of spectra is the same and is not enough to perform spectrophotometry. Indeed, any deviation from the real shape of the spectrum is carried over to the other exposures through the factor  $\mathcal{A}_\mathrm{templ}(\lambda_n)$.\\
Now that the shape of the spectra is adjusted, it is possible to remove the overall flux level fluctuations $\mathcal{B}(t_i)$. This is done by rescaling each spectrum order by order to its average:
\begin{equation}
\label{Eq:color_corrected_rescaled_spectra}
\left[\mathrm{C_{cc,\,r}}(\lambda_n, t_i)\right]_m=\dfrac{\left[\mathrm{C_{cc}}(\lambda_n, t_i) \right]_m}{ \langle \mathrm{C_{cc}}(\lambda_n, t_i)\rangle_m}=\dfrac{\left[\mathrm{F}(\lambda_n, t_i)\right]_m}{\langle \mathrm{F}(\lambda_n, t_i)\rangle_m}\,,
\end{equation} 
where we have used the independence of $\mathcal{B}(t_i)$ from wavelength, and assumed that $\mathcal{A}_\mathrm{templ}(\lambda_n)$ can be brought out of the average within order $m$. At this point, the spectra have the same level in the continuum and can be combined.\\
Now, recall that $\mathrm{F} \propto R_p^2\cdot\mathrm{F_p}(\lambda_n, t_i) +  R_\star^2\cdot\mathrm{F_\star}(\lambda_n)$. While the star is assumed to be stable over the course of an observation, the planetary spectral lines move as a result of its Doppler motion, hence its time dependence. Our aim is now to remove $R_\star^2\cdot \mathrm{F_\star}$ to isolate the planet signal. This is done by building a master spectrum $\mathrm{M}_\star$ containing only the stellar spectrum and the planetary continuum, and normalizing each exposure by the master spectrum. As common in the literature, we obtain the master spectrum with a median in time of the color corrected, rescaled spectra Eq. \ref{Eq:color_corrected_rescaled_spectra}. Since the planet moves in time by about 0.5 -- 3.5 pixels per exposure, its lines are mostly removed from the master. By splitting the planet flux in its line and continuum contribution ($\mathrm{F}_\mathrm{p,\,lines}$ and $\mathrm{F}_\mathrm{p,\,cont}$):
\begin{equation}
\left[\mathrm{M}\right]_m=\mathrm{med_t}\left[ \mathrm{C_{cc,\,r}}(\lambda_n, t_i) \right]_m \approx \dfrac{\left[\mathrm{F}_\star(\lambda_n) + \mathrm{F}_\mathrm{p,\,cont}(\lambda_n)\right]_m}{\langle \mathrm{F}_\star(\lambda_n) + \mathrm{F}_\mathrm{p,\,cont}(\lambda_n) \rangle_m }\,,
\end{equation}
where we neglected the flux contained in the planetary spectral lines at the denominator. Finally, by dividing $\mathrm{C_{cc,\,r}}(\lambda_n, t_i)$ by the master spectrum, we obtain:
\begin{equation}
\label{Eq:planet_excess}
\left[ \mathrm{C_{cc,\,r,\,tn}}(\lambda_n, t_i)\right]_m = \dfrac{\left[\mathrm{C_{cc,\,r,}}(\lambda_n, t_i)\right]_m}{\left[\mathrm{M}(\lambda_n)\right]_m} = \dfrac{R_\mathrm{p}^2\cdot \left[ \mathrm{F_{lines\,p}}(\lambda_n, t_i)\right]_m}{\left[R_\star^2\cdot \mathrm{F_\star}(\lambda_n) + R_\mathrm{p}^2 \cdot \mathrm{F}_\mathrm{p,\,cont}(\lambda_n) \right]_m}+1\,.
\end{equation}
What we measure, is thus the planetary lines normalized to the stellar plus planetary continuum.\\
Finally, we applied a high-pass filter by computing the standard deviation of each pixel in time (i.e. across the full spectral sequence) and applying a threshold 3 times above the median level of the noise (the exact choice for the threshold level does not influence the final result). For each exposure and each order, we fitted a second-order polynomial to the spectra after rejecting strong outliers and masked pixels. We then divided the data by the fitted polynomial. Eventually, we applied the cross-correlation function.\\
The planet continuum itself can not be recovered. Indeed, the rescaling in Eq. \ref{Eq:color_corrected_rescaled_spectra} must be carried out order by order, because within one order $\mathcal{A}(\lambda_n, t_i)$ is approximately constant. The same holds for the planetary continuum, which is thus removed from our analysis as a by-product. Alternative approaches use a polynomial normalization, with the same outcome. Recently, \cite{Cauley2019} claimed that they perform flux calibration on Echelle spectra similar to ours. Such an approach has a potentially enormous impact on the study of exoplanet atmospheres with this technique, because it would preserve the planetary continuum, which would already be detectable with currently achieved precisions \citep{Pino2018b}.\\

\section{Line weighted, binary mask cross-correlation function}
\label{sec:appendix_CCF}
Functionally, this CCF is a weighted average of a wavelength dependent signal $S(\lambda)$, in our case the planetary spectrum normalized to the continuum (Sec. \ref{sec:appendix_prepararion_of_spectra}), on the spectral lines considered in the mask,
\begin{equation}
\label{Eq:CCF_complete}
\mathrm{CCF}(v)=\dfrac{\sum^\mathrm{orders} w_\mathrm{order} \sum_{i=1}^{N_\mathrm{lines}} \int_\mathrm{order}  S(\lambda)\cdot \left.M_i\left(\lambda\right)\right|_v\cdot w_i d\lambda}{\sum^\mathrm{orders} w_\mathrm{order} \sum_{i=1}^{N_\mathrm{lines}}\int_\mathrm{order} \left.M_i\left(\lambda\right)\right|_v w_i}\,.
\end{equation}
Within each order, to each of the $N$ lines considered, we associate a binary mask $M_i$ that has a value of 1 within a waveband $0.82~\mathrm{km~sec^{-1}}$ wide (1 HARPS-N nominal pixel) around each considered line shifted to account for a radial velocity $v$, 0 elsewhere. Each order is weighted according to the signal-to-noise ratio of the observations ($w_\mathrm{order} = 1/\sigma_\mathrm{order}$, where $\sigma_\mathrm{order}$ is the photometric dispersion of the order computed between pixels 1024 and 3072, and only orders with signal-to-noise ratio larger than 35 were kept), and each line is weighted ($w_i$) according to its information content. In the case of the G2 mask that we used, this is the contrast of the spectral line, but different applications may require different weighting schemes. Since the width of the masks in the wavelength space changes with radial velocity, it is important to compute the normalization at every value of $v$. \\
Computationally, it is convenient to recast Eq. \ref{Eq:CCF_complete} to have an effective weight for every pixel in the detector. Practically, each binary mask can span one or more complete pixels and fractions of pixels at the edges. For a single line $i$, we can expand the integral by co-adding contributions from each pixel or pixel part that falls within the binary mask $M_i$. If we label each pixel by $j$, and call $\Delta\lambda_j$ the width of the pixel in wavelength space, then pixels entirely within the mask contribute to the spectrum with $\overline{\Delta\lambda_j} = \Delta\lambda_j$, while pixels at the edges of the mask contribute with $\overline{\Delta\lambda_j} < \Delta\lambda_j$. Thus:
\begin{equation}
\label{Eq:CCF_discrete}
\mathrm{CCF}(v)=\dfrac{\sum^\mathrm{orders} \sum_{i=1}^{N_\mathrm{lines}} \sum_{j=1}^{N_\mathrm{pixels\,in}\,\left. M_i\right|_v} S(\lambda_j) \cdot \left(   w_\mathrm{order}  \cdot w_i \cdot \overline{\Delta \lambda_j} \right)}{\sum^\mathrm{orders}\sum_{i=1}^N \sum_{j=1}^{N_\mathrm{pixels\,in}\,\left. M_i\right|_v} \left(w_\mathrm{order}  \cdot  w_i \cdot \overline{\Delta \lambda_j} \right)}\,.
\end{equation}
The term in parenthesis is the effective weight for each pixel in each order, and is a unique property of each mask considered. Written in this form, the calculation can be conveniently performed using matrix calculation.\\
We computed the CCF in each order of each exposure by sliding the binary mask between $-400~\mathrm{km~s^{-1}}$ and $400~\mathrm{km~s^{-1}}$ in steps of $2.7~\mathrm{km~s^{-1}}$ (1 nominal HARPS-N resolution element, containing about 3 nominal HARPS-N pixels). With this choice, we were entitled to treat each CCF point as statistically independent from the others, since their information comes from separate resolution elements. For each exposure, we then obtained a total CCF by summing the CCFs of each single order. With a similar procedure, we computed the normalization at the denominator in Eq. \ref{Eq:CCF_discrete}.\\
The peak of the CCF is found at a different position in every exposure, due to the planet motion around its host. The juxtaposition of all exposures provides a planet trace. We then assumed a circular orbit for the planet and shift the CCF in each exposure for different values of the tangential velocity of the planet $K_\mathrm{p}$. For every combination, we interpolated the total CCFs in each exposure to a common velocity grid, and summed them. The resulting 1D CCF is maximized when the individual exposures are correctly aligned in the rest-frame of the planet.

\section{Radiative transfer code}
\label{sec:appendix_model}
We solved the radiative transfer equation in its integral form, employing a ``linear in optical depth'' approximation for the source function, which is valid for a non-scattering atmosphere \citep{Toon1989}. We employed 200 logarithmically spaced layers between $10^5~\mathrm{bar}$ and $10^{-12}~\mathrm{bar}$, covering the full region where lines are generated with enough spatial resolution. This was verified with a step doubling procedure.\\
For a given temperature-pressure profile, we assumed equilibrium chemistry and calculated volume mixing ratios using the publicly available \texttt{FastChem} code  version 2 (\citealt{Stock2018}; Stock, Kitzmann \& Patzer in prep.). Our opacities are calculated by employing the VALD3 database \citep{Piskunov1995, Ryabchikova1997, Kupka1999, Kupka2000, Ryabchikova2015, K14, BKK, BK, BPM, BWL, FMW, K10, BLNP, NWL, LGWSC, PTP, MFW, BHN, RHL, WLSC, K13, BA-J, KK, PGHcor, BSScor, BGHR, HLGN, B, T83av, RU}. While the VALD3 database offers line lists for a variety of atomic and molecular species, we limited this study to \ion{Fe}{1}, \ion{Fe}{2}, \ion{Ti}{1}, \ion{Ti}{2}, expected to be the most spectrally active species in KELT-9b \citep{Hoeijmakers2018_k9}. We computed opacity tables by broadening the lines with a Voigt profile accounting for thermal and natural broadening, and we used partition functions by \cite{Barklem2016} to obtain opacities as a function of temperature, over a fine grid in wavelength ($\Delta \lambda = 0.001~\mathrm{\AA}$) over the full HARPS-N range. At this resolution, the single lines in the atmosphere are resolved by a factor of 20 to 30, making our code effectively line-by-line. Our $\mathrm{H}^-$ bound-free opacity comes from \cite{John1988}, in particular their Eq. (4). We also note a possible imprecision in the units for $\lambda_0$ and $\alpha = hc/k_b$ in that paper, which appear to be inconsistent. If $\lambda_0$ is taken in $\mu\mathrm{m}$ as the author suggests, the correct value for $\alpha$ to insert in Eq. (3) is $1.439\cdot10^4$ rather than $1.439\cdot10^8$.\\
We validated our code by reproducing the position and depth of iron lines in a $\log g = 4.5$, $T_\mathrm{eff}=4,500~\mathrm{K}$ PHOENIX model \citep{Husser2013}, adopting the temperature profile provided in the `ATMOS.fits' file. For such a star, iron lines are modelled in LTE, which we also assumed. We did not attempt to reproduce the pressure broadened wings and micro-turbulence broadening in the stellar spectrum, because HARPS-N is only sensitive to the core of the planetary iron lines and micro-turbulence is degenerate with rotational broadening at our level of precision. We also validated the continuum in our model by reproducing it with petitRADTRANS \citep{Molliere2019}, finding agreement to within a few percent over the HARPS-N range.

\section{Significance of the detection, model comparison and confidence intervals}
\label{sec:appendix_basic_statistics}
This Appendix presents practical details on how we treated $\log \mathcal{L}$ to (1) assess the significance of our detection, (2) perform model comparison, (3) extract confidence intervals. All of these tasks can be performed using Wilk's theorem \citep{Wilks1938}. An extensive literature on the topic is available (e.g. \citealt{Lampton1976} treats most of these problems in a very clear manner), and we specialize the discussion to our method. We also provide a practical method to marginalize the likelihood distribution.\\
Given a model $S$ with $p$ parameters, a model $S_\mathrm{nested}$ is nested to it if it can be obtained from $S$ by fixing $q<p$ parameters. In this case, $\max \mathcal{L(S)} \geq \max \mathcal{L(S_\mathrm{nested})}$. Wilk's theorem states that the likelihood ratio test (LRT) metric
\begin{equation}
\label{Eq:LRT}
\mathrm{LRT} = -2\ln \dfrac{\max \mathcal{L(S_\mathrm{nested})}}{\max \mathcal{L(S)}} = -2 \ln\left[ \max \mathcal{L(S_\mathrm{nested})} - \max \mathcal{L(S)} \right]
\end{equation}
is distributed as a $\chi^2$ distribution with $q$ degrees of freedom under the null hypothesis that $S_\mathrm{nested}$ is true.\\
The application to model comparison is straightforward: in our case, $S_{\mathrm{Fe}}$ and $S_{\mathrm{Ti}}$ are nested in  $S_{\mathrm{Fe,\,Ti}}$, and $q=1$. The survival function of a $\chi^2$ distribution with 1 degree of freedom evaluated in LRT gives the probability that the measured LRT difference would be observed by chance alone. A high probability indicates that the null hypothesis that the nested model is true can be excluded. We convert this probability to $\sigma$ values using a two tailed Gaussian distribution.\\
To assess the significance of the detection, we created a nested model with $\mathrm{VMR_{Fe}}=0$. We then compared this to our preferred model $S_{\mathrm{Fe}}$. Using the properties of \ref{Eq:LRT}, we computed the probability at which the null hypothesis that a model without any spectral line can be excluded (again, $q=1$).\\
It is maybe less evident that the same theorem allows us to compute confidence intervals. A clear explanation is found in \cite{Cash1976}, which we summarize. Assume that we are interested in the confidence interval on parameter $\theta$, which can have values ($\overline{\theta}_1$, $\overline{\theta}_2$, $\dots$). First, we fix $\theta = \overline{\theta}_1$, and look for the maximum likelihood by varying the rest of the parameters. Practically, this is a nested model with $q=1$. Thus, we can apply Wilk's theorem to compute the probability that the null hypothesis that $\theta = \overline{\theta}_1$ is excluded. We then move to the next value of $\theta$, and repeat the operation. The locus of $\theta$ values for which we obtain probabilities lower than a threshold $\alpha$ gives the corresponding confidence interval.\\
Given a sufficiently fine grid of likelihoods, another equivalent option is to directly marginalize the likelihood. However, in general, $\exp \left(\log \mathcal{L} \right)$ can be a computationally untreatable number. We thus normalize the likelihood to its maximum prior to exponentiating, by computing
\begin{equation}
\overline{\mathcal{L}} = \exp \left[\log \mathcal{L} - \max \log \mathcal{L}\right]\,.
\end{equation}
This quantity can then be marginalized, and correctly normalized a-posteriori. The contour levels can be obtained as percentiles of the resulting marginalized distribution.

\section{Cross-correlation to likelihood mapping by to Brogi \& Line (2019)}
\label{sec:appendix_test_retrieval}
To check the consistency of our method, we retrieved $v_\mathrm{sys}$ and $K_\mathrm{p}$ using the framework described in \cite{Brogi2019}, and the best-fitting model computed and scaled as explained in Sec. \ref{sec:methods_models} and Appendix \ref{sec:appendix_pipeline}. In this scheme, the cross covariance $R$ between data and the best-fitting model (rather than a binary mask) is computed. As such, the retrieval is model-dependent in line with its main application to determine atmospheric properties of exoplanets. Cross-covariance values are translated into log-likelihood via the formula:
\begin{equation}
\label{Eq:likelihood_Brogi}
\log(L) = -\frac{N}{2} \log[s_f^2 + s_g^2 - 2R]\,,
\end{equation}
where $s_f$ and $s_g$ are the data and model variance, respectively. A Markov Chain Monte Carlo is driven by the likelihood in Eq. \ref{Eq:likelihood_Brogi}, and run via the Python package $\tt{emcee}$. Confidence intervals are determined by marginalising the posterior distributions and computing confidence intervals consistently with the line-weighted binary mask method (see Sec. \ref{sec:appendix_basic_statistics}).\\
We compared the likelihood distributions for $K_\mathrm{p}$ and $v_\mathrm{sys}$ obtained with the two methods in Fig. \ref{Fig:likelihood_velocities}. The frameworks give results consistent at $1\sigma$. The significance of the detection with the framework by \cite{Brogi2019} is $10.3\sigma$, which is higher than the significance found with the line weighted binary mask case, and the confidence intervals on $K_\mathrm{p}$ and $v_\mathrm{sys}$ are consequently tighter. Possible explanations include (1) the larger amount of pixels and line shape information used in the \cite{Brogi2019} case, (2) the fact that, in the line weighted binary mask approach, we do not use a mask tailored to the planetary spectrum but rather a G2 stellar spectrum. A more detailed comparison will be target of dedicated work.

\bibliographystyle{aasjournal}
\bibliography{My_biblio}

\acknowledgements
LP acknowledges helpful discussions with the exoplanet atmosphere team at Anton Pannekoek institute for astronomy, particularly K. Todorov and J. Arcangeli, and with Phil Uttley. We thank the anonymous referee, who's insightful feedback improved the quality and clarity of the manuscript. LP thanks P. Molli\`ere for the prompt support with the validation of the radiative transfer code. The research leading to these results has received funding from the European Research Council (ERC) under the European Union’s Horizon 2020 research and innovation programme (grant agreement no. 679633; Exo-Atmos). Based on observations made with the Italian {\it Telescopio Nazionale Galileo (TNG)} operated on the island of La Palma by the {\it Fundaci\`on Galileo Galilei} of the INAF at the  {\it Observatorio Roque de los Muchachos}. This work has made use of the VALD database, operated at Uppsala University, the Institute of Astronomy RAS in Moscow, and the University of Vienna. This research made use of Astropy,\footnote{http://www.astropy.org} a community-developed core Python package for Astronomy \citep{astropy2013, astropy2018}. J.M.D acknowledges support by the Amsterdam Academic Alliance (AAA) Program. M.B. acknowledges support from the UK Science and Technology Facilities Council (STFC) research grant ST/S000631/1. LM acknowledges support from the University of Rome Tor Vergata through ‘Mission: Sustainability 2017’ fund. GS acknowledges financial support from ‘Accordo ASI–INAF ’ No. 2013-016-R.0 2013 July 9 and 2015 July 9. R.A. acknowledges the financial support of the SNSF. We acknowledge support from the INAF/FRONTIERA project through the “Progetti Premiali" funding scheme of the Italian Ministry of Education, University, and Research.

\end{document}